\documentclass[epjCONF]{svjour}
\usepackage{graphics}
\usepackage[varg]{txfonts} 
\usepackage[latin1]{inputenc}
\newcommand{\intd}{{\textrm{d}}}
\newcommand{\dslash}{\partial\hspace{-1.85mm}/}

\newcommand{\intddd}[1]{\int \frac{\intd^3 #1}{(2\pi)^3}}

\newcommand{\beq}{\begin{equation}}
\newcommand{\eeq}{\end{equation}}
\newcommand{\bea}{\begin{eqnarray}}
\newcommand{\eea}{\end{eqnarray}}
\newcommand{\bce}{\begin{center}}
\newcommand{\ece}{\end{center}}

\newcommand{\pkt}{\;.}

\newcommand{\inv}[1]{\frac{1}{#1}}

\newcommand{\Tr}{{\rm Tr}}

\newcommand{\ave}[1]{\left<#1\right>}

\session-title{Hot and Cold Baryonic Matter -- HCBM 2010}
\begin{document}
\title{Recent Theoretical Developments in the QCD Phase Diagram}
\author{Jochen Wambach\inst{1}\fnmsep\inst{2}\fnmsep\thanks{\email{wambach@physik.tu-darmstadt.de}}}
\institute{Institut f\"ur Kernphysik\\
Technische Universit\"at Darmstadt\\
64289 Darmstadt\and 
GSI Helmholtzzentrum f\"ur Schwerionenforschung GmbH\\
64291 Darmstadt
}
\abstract{
In this talk I discuss three recent developments in the theoretical understanding of the phase diagram of the strong interaction. The first topic deals with the comparison of model calculations of the quark-hadron transition at vanishing quark chemical potential with state-of-the-art lattice QCD results. In the second relates to the size of a possible 'quarkyonic phase'. The third deals with the occurence of inhomogeneous chiral phases.
} 
\maketitle
\section{Introduction}
\label{intro}
Exploring the chiral and deconfining properties of strong-interaction
matter at high temperatures and large densities is one of the central themes in 
nuclear- and astrophysics \cite{PBMW,Alford}.  The general phase structure and, in particular, the possible 
existence of a (chiral) critical endpoint (CEP) for finite net baryon number density 
and its consequences for the phase structure of QCD at lower temperatures are
much under debate. High-energy heavy-ion collision experiments at RHIC
and the SPS have started to look for experimental evidence of the CEP and
experiments at future heavy-ion facilities (FAIR and NICA), have been
designed to probe the relevant high-density region in the QCD phase
diagram.

Gaining insight into the properties of QCD matter at non-vanishing 
quark chemical potential ($\mu$) is thus of great interest.
However, the relevant region of the phase diagram is not easy to access 
directly in QCD. At non-zero values of $\mu$ the notorious fermion
sign problem prohibits straightforward lattice QCD simulations. Thus one 
currently resorts to models that capture some of the essential features
of QCD, such as the basic symmetries and their breaking patterns. 
Although limited in their predictive power, such models, nontheless, give
valuable insight into various physical effects governing the phase structure,
in particular the existence of a chiral CEP, a confined but chirally restored 
phase at low temperatures ('quarkyonic matter') and/or the occurence of
inhomogeneous chiral phases.

Model predictions for the equation of state (EoS) of QCD matter can be compared
with lattice simulations at vanishing $\mu$. Due to large lattice sizes,
small quark masses and algorithmic improvements, the latter have become very 
accurate recently. Such comparisons will be discussed in Sect. 2. Based on large $N_c$
arguments for the $T$- and $\mu$-dependence of the pressure, a 'quarkyonic'
phase has been conjectured at low $T$ and large $\mu$. In this phase, quarks 
and gluons are still confined in hadrons (mostly baryons) but chiral symmetry is 
restored, leading to parity degeneracy in the excitation spectrum. There has been much 
debate recently about the size and the detailed structure of this phase. Theoretical 
arguments and empirical constraints will be discussed in Sect. 3. An exciting possibility 
is the occurence of inhomogeneous chiral phases at low $T$ and large $\mu$, possibly 
entangled with inhomogeneous color superconducting phases. Inhomogeneous phases 
are predicted in $1+1$ dimensional models of QCD in the large $N$ limit and their 
three dimensional analogue could shed light on the nature of the quarkyonic phase. Recent 
results and preliminary conclusions will be discussed in Sect. 4.   

\section{The Quark-Hadron Transition at small $\mu$}
\label{sec:1}

I begin with the behavior of strong-interaction matter at very small $\mu$ and
large $T$, as it is encountered in the early universe. Modern models such as
the PNJL or the PQM model incorporate (approximate) chiral symmerty and 
its spontaneous breaking in the vacuum, the axial $U_a(1)$ ano- maly and the 
heavy quark limit in terms of the temporal Polyakov loop. 

\subsection{The Polyakov Quark Meson Model}
As an example I will 
discuss the EoS predictions of the PQM model (the results of the PNJL model are
very similar). The 3-flavor PQM model is basically the linear sigma model with
up, down and strange quarks coupled the flavor octet of scalar and 
pseudoscalar meson fields $\sigma_a$ and $\pi_a$:
\beq
{\cal L}_{quark} = \bar q\left(i \dslash-G\frac{\lambda_a}{2}
\left(\sigma_a+
i\gamma_5\pi_a\right)\right)q
\eeq
and the mesonic Lagrangian given by 
\bea
&{\cal L}_{meson}=\Tr(\partial_\mu M^\dagger\partial^\mu M)-m^2\Tr(M^\dagger M)
+\Tr[H(M+M^\dagger)]
\nonumber\\
&-\lambda_1[\Tr(M^\dagger M)]^2-\lambda_2\Tr(M^\dagger M)^2
+c\left(\det(M)+\det(M^\dagger)\right)
\nonumber\\
\eea
where
\beq
M=\sum_a\frac{\lambda_a}{2}\left(\sigma_a+i\pi_a\right);\;H=\sum_a\frac{\lambda_a}{2}h_a\pkt 
\eeq
The term involving the determinant of $M$ and $M^\dagger$ incorporates the axial anomaly.
This is supplemented by the Polyakov loop expectation value
\beq 
\ell=\inv{N_c}\Tr P\exp\left[i\!\!\int_0^\beta 
\!\!d\tau\, A_4({\bf x,\tau})\right]
\eeq
and the covariant derivative of the temporal
gauge field $A_4({\bf x,\tau})$ as well as a potential term such that 
\beq
{\cal L}_{pol}=-\bar q\gamma_4A_4q-{\cal U}(\ell,\bar\ell)\pkt 
\eeq
The total PQM Lagrangian then reads  
\beq
{\cal L}_{PQM}={\cal L}_{quark}+{\cal L}_{meson}+{\cal L}_{pol}\pkt
\eeq
Using the mean-field approximation, the parameters are adjusted to
the vacuum meson masses and their weak decay constants. The Polyakov
potential ${\cal U}(\ell,\bar\ell)$, on the other hand, is determined from a fit
to the pure gauge lattice EoS. There are various choices for ${\cal U}$ in the 
literature such as a polynomial in $\ell$ and $\bar\ell$  based on 
Landau-Ginzburg theory \cite{Ratti}, a logarithmic
ansatz motivated by the Haar measure of the $SU(3)$ color gauge group 
\cite{Roesner} or a choice derived from the strong-coupling expansion of QCD, 
proposed by K. Fukushima \cite{FukuPot}.
\subsection{Mean-Field Results}

\begin{figure}[h]
\resizebox{1.0\columnwidth}{!}{%
\includegraphics{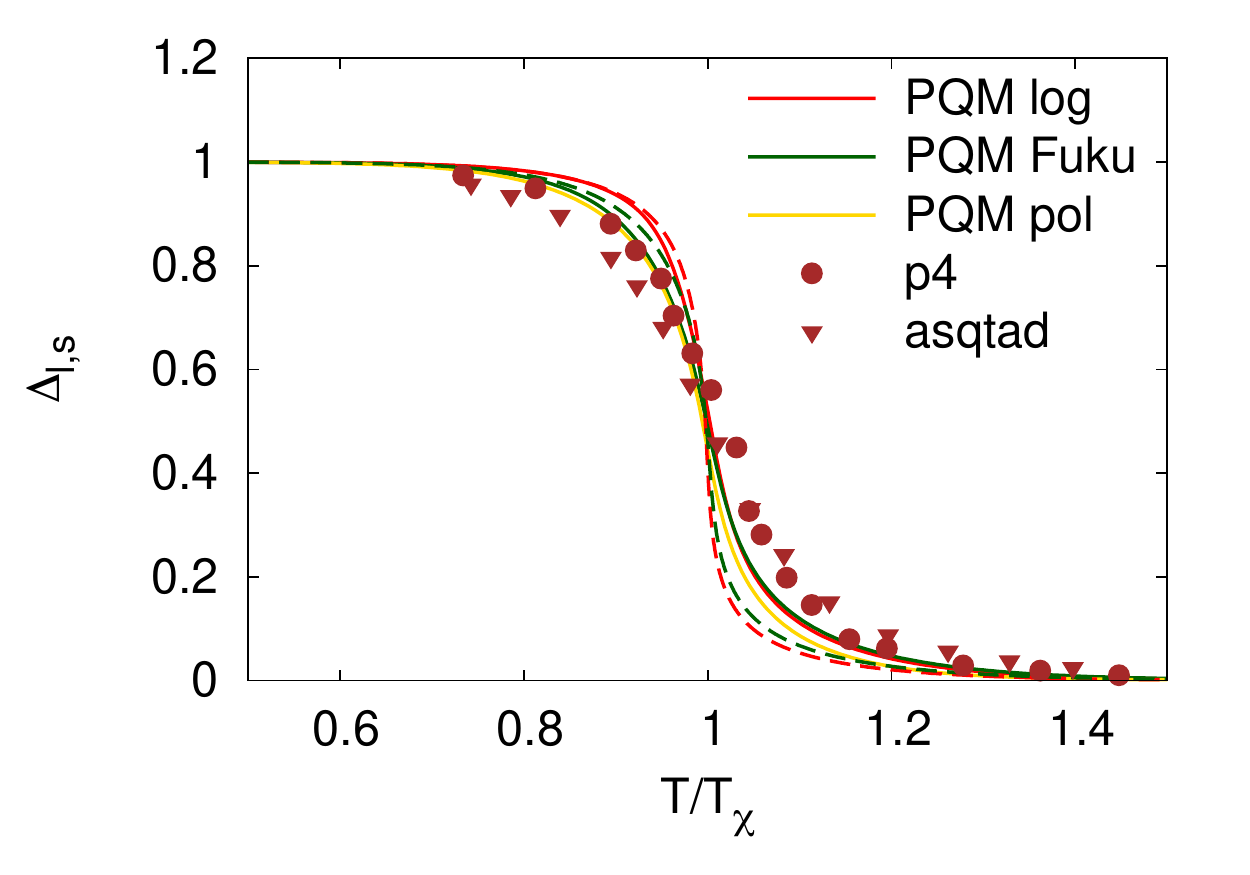} }
\resizebox{1.0\columnwidth}{!}{%
\includegraphics{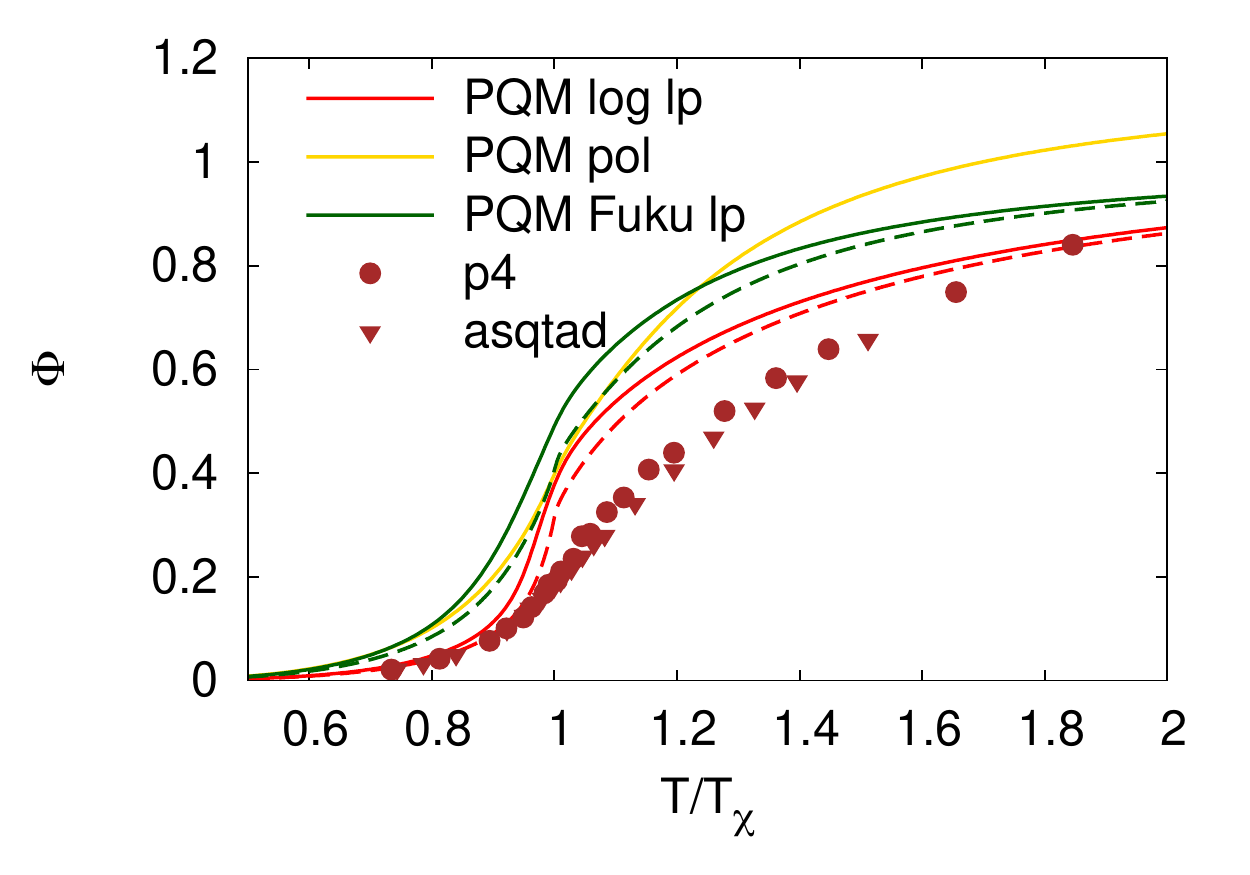}}
\caption{The normalized quark condensate $\Delta_{l,s}$ (upper part) and the Polyakov 
loop expectation value (lower part) as a function of $T/T_\chi$ at vanishing $\mu$. Here $T_\chi$ denotes 
the (pseudo)critical chiral transition temperature. The lattice data are taken from Ref. \cite{HOTQCD,WB}}
\label{fig:MF-latt}       
\end{figure}

At finite $T$ and $\mu$ one  evaluates the mean-field grand potential
as 
\beq
  \Omega(T,\mu;\sigma_l,\sigma_s,\ell,\bar\ell)= U \left(\sigma_{l},\sigma_{s}\right) +
  \Omega_{\bar{q}{q}} \left(\sigma_{l},\sigma_{s}, \ell,\bar\ell\right) +
  {\cal U}\left(\ell,\bar\ell\right)\\
 \eeq
 where $U \left(\sigma_{l},\sigma_{s}\right)$ denotes the effective meson potential in
 terms of the light quark condensate $\sigma_l=\langle \bar{l}l\rangle$ and the strange quark 
 condensate $\sigma_s=\langle \bar{s}s\rangle$. The fermionic part $\Omega_{\bar{q}{q}} $ 
 involves both quark condensates and the Polyakov loop expectation value and is given by 
 \bea
 \Omega_{\bar qq}=
     -2N_fT\int\frac{d^3p}{(2\pi)^3}\left\{\ln g(T,\mu)+\ln g(T,-\mu)\right\}
 \eea
 with
 \bea
 &g(T,\mu)=\left[1 + 3 \ell e^{-(E_{p}-\mu)/T}
     +3 \bar\ell e^{-2(E_{p}-\mu)/T}+
        e^{-3(E_{p}-\mu)/T}\right]
        \nonumber\\
 \label{gtm}
 \eea
where $E_p=\sqrt{\vec p^2+M_q^2}$ denotes the quark quasi-particle energy with constituent mass
$M_q$. Eq. \ref{gtm} shows that in the hadronic phase single- diquark contibutions are suppressed, since 
in the confined region both $\ell$ and $\bar \ell$ vanish. Finally the phase diagram is determined
via the stationarity condition:
\beq
\left.\frac{ \partial \Omega}{\partial
      \sigma_l}= \frac{ \partial \Omega}{\partial \sigma_s}  = \frac{ \partial \Omega}{\partial \ell}  =\frac{ \partial \Omega}{\partial \bar\ell}
  \right|_{\rm{min}} = 0\pkt
\eeq
In Fig. \ref{fig:MF-latt} predictions of a PQM mean field calculation \cite{SWW} are compared with recent lattice data of the HOTQCD and the WB colaborations \cite{HOTQCD,WB}. 
While the general agreement is quite good, the quark-hadron transition is predicted to be sharper than in the lattice simulations (especially for physical bare quark masses \cite{WB}). The same also holds for the pressure and other bulk thermodynamic quantities. We will return to this point in the next section.
\subsection{Including Fluctuations}
\begin{figure}[h]
\resizebox{1.0\columnwidth}{!}{%
\includegraphics{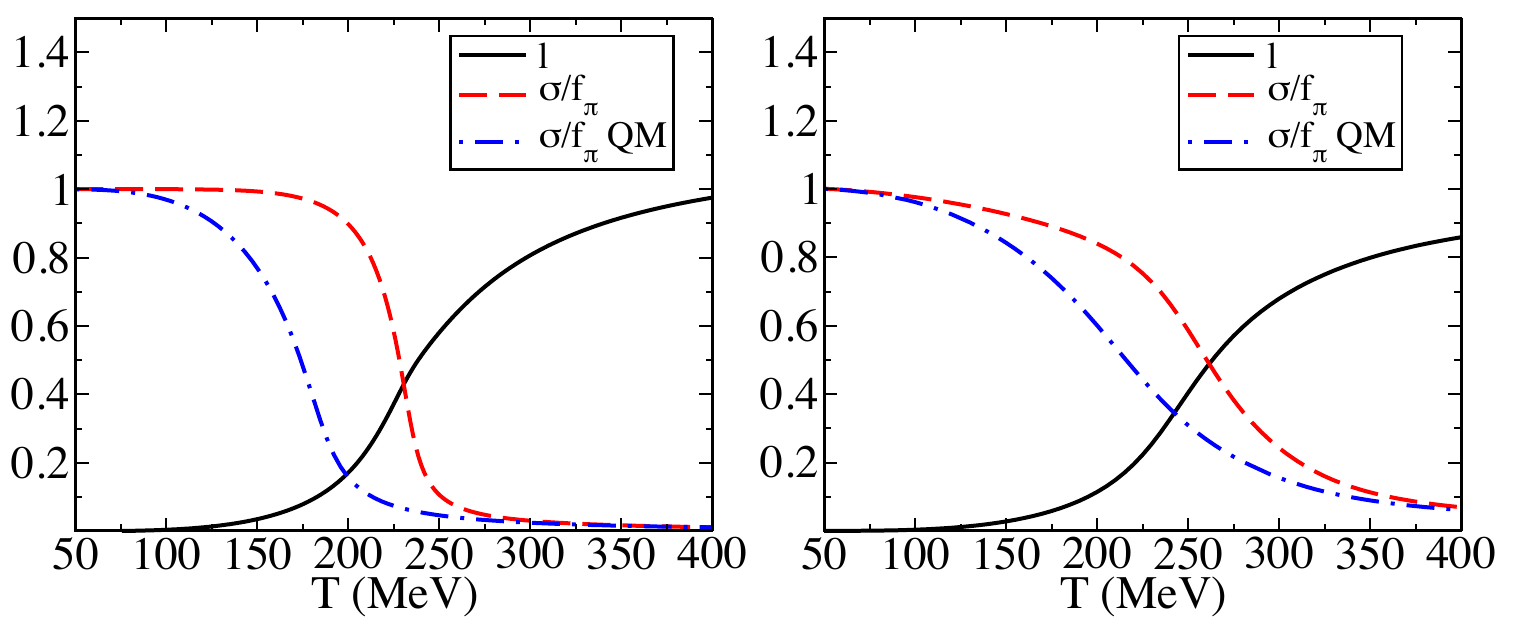} }
\caption{A comparison between mean-field results of the $N_f=2$ PQM model (left part) and those from the FRG (right part) for the chiral condensate (red and blue lines) and the Polyakov-loop expectation value (full lines) \cite{SSFR}. The curves marked in blue display pure QM results.}
\label{fig:FRG}       
\end{figure}

The relative sharpness of the quark-hadron transition in mean-field theory can be 
understood as an effect of the omission of quantum fluctuations. These can be included
most efficiently in the functional renormalization group approach (FRG). The FRG is based on an
infrared regularization of the grand potential, which then becomes a momentum scale 
dependent quantity $\Omega_k$ with a scale parameter $k$. The full potential, including all
quantum fluctuations is obtained in the limit $k\to 0$, i.e.
\beq
\Omega(T,\mu)=\lim_{k\to 0}\Omega_k(T,\mu)\pkt
\eeq 
The momentum flow equations for the QM model have been derived in \cite{SWa} and 
extended to include the Polaykov loop effects as a background field in \cite{SSFR}. The resulting flow equations for $N_f=2$ read
\bea
\partial_k\Omega_k&=&\frac{k^4}{12\pi^2}\left[\frac{3}{E_\pi}(1+2n_B(E_\pi))
+\frac{1}{E_\sigma}(1+2n_B(E_\sigma))\right .\nonumber\\
&-&\left .\frac{N_cN_f}{E_q}\left(1-n_q(\ell,\bar\ell)-n_{\bar q}(\ell,\bar\ell)\right)\right]\pkt
\label{PQMflow}
\eea
Here $n_B$ are the bosonic distribution functions with the pion and sigma energy
\beq
E_\pi^2=1+2\Omega'_k/k^2\quad E_\sigma^2=1+2\Omega'_k/k^2+4\phi^2\Omega''_k/k^2
\eeq
where the prime denotes a functional derivative with respect to the chiral condensate $\phi=\ave{\sigma}$,
i.e.
\beq
\Omega'_k=\partial\Omega_k/\partial\phi\quad etc\quad \phi=\ave{\sigma}\pkt
\eeq 
The explicit form of the fermionic distribution functions in the Polyakov loop background field $n_q(\ell,\bar\ell)$ and $n_{\bar q}(\ell,\bar\ell)$ are given in \cite{SSFR} and involve the quark quasi-particle energy 
\beq
E_q^2=1+G\phi^2/k^2
\eeq
In the absence of the background gluon field they reduce to the Fermi-Dirac distribution for constituent quarks, i.e. expression (\ref{PQMflow}) coincides with that in \cite{SWa}. As can be seen from Fig. \ref{fig:FRG} the pure mean-field transition (left part) is significally softened when fluctuations are included via the FRG. 

\begin{figure}[h]
\resizebox{0.9\columnwidth}{!}{%
\includegraphics{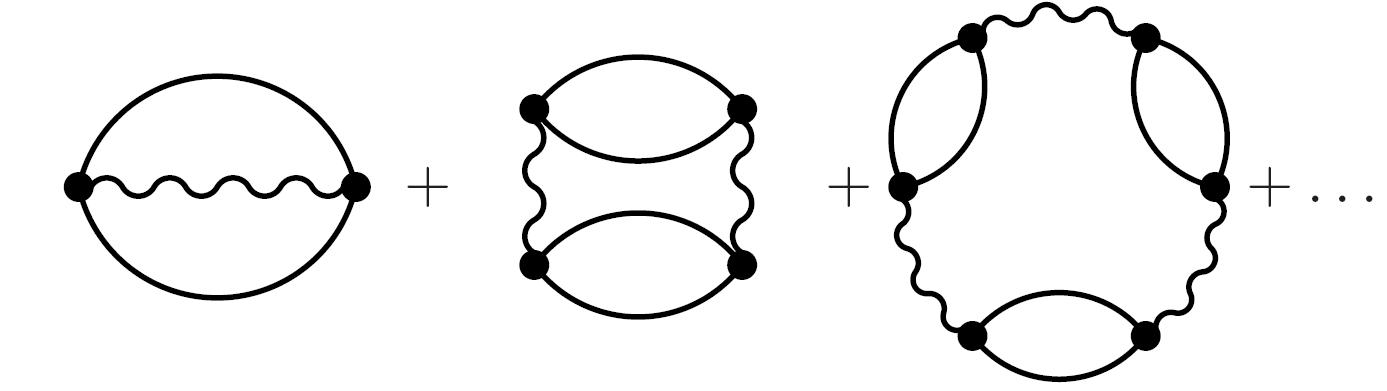} }
\caption{Next-to-leading-order $1/N_c$-corrections to the thermodynamic potential in the (P)NJL model. The wiggly lines denote the four-quark contact interaction of the model. These ring diagrams can be summed to all orders.}
\label{fig:ringsum}      
\end{figure}

\begin{figure}[h]
\resizebox{1.0\columnwidth}{!}{%
\includegraphics{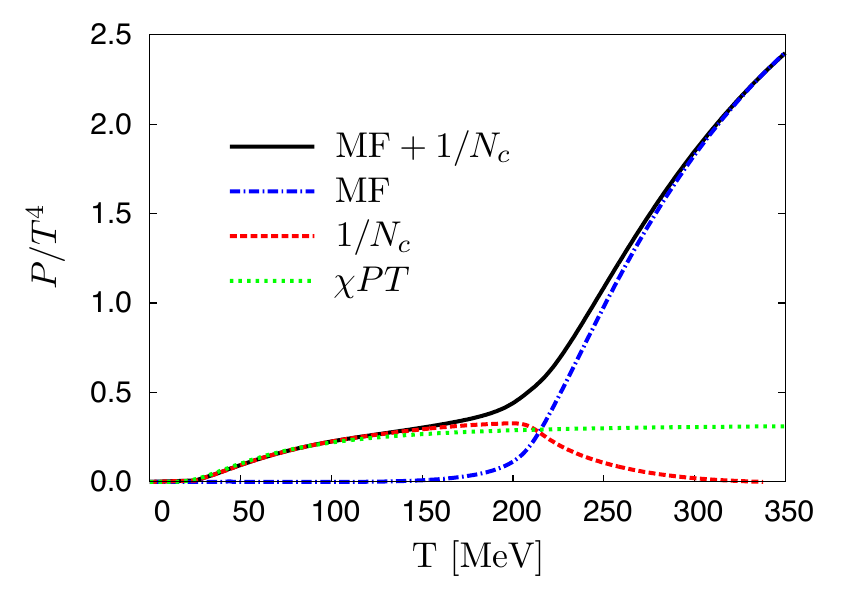} }
\caption{Various contributions to the normalized pressure $P/T^4$ in the three-flavor PNJL model \cite{Radza}. The dashed-dotted line indicates the mean-field result, while the dashed line displays the purely mesonic contributions (mostly pions). Up to temperatures of $\sim$ 150 MeV these agree well with the results from chiral perturbation theory \cite{GaLeu} (dotted line).  
}
\label{fig:pfluct}      
\end{figure}

\noindent
This softening can be understood physically by considering the next-to-leading-order $1/N_c$-corrections to the mean field result for $\Omega$ \cite{Oertel}, i.e.
\beq
\Omega=\Omega_{MF}+\delta\Omega\pkt
\eeq
Diagrammatically this amounts to a summation of all ring diagrams (Fig. \ref{fig:ringsum})
and represent mesonic contributions to $\Omega$ in leading order:
\beq
\delta\Omega=\sum_M\Omega_M\pkt
\eeq
In imaginary time $\Omega_M$ is represented by a Matsubara sum involving the mesonic polarization 
function $\Pi_M$ 
\beq
\Omega_M=\intddd{q}\frac{T}{2}\sum_{i\omega_q}
\ln(1-2G\Pi_M(i\omega_q,\vec q))\nonumber
\eeq
which can be Wick-rotated and leads to
\beq
\Omega_M=-\intddd{q}\int_0^\infty\frac{d\omega}{\pi}\left(1+2n_B(\omega)\right)\phi_M
\eeq
where $n_B$ is the finite temperature bosonic occupation probability and $\phi_M$ can be interpreted as an 'in-medium' phase shift in the Beth-Uhlenbeck sense. Explicitly one has
\beq
\phi_M=\frac{1}{2i}\ln\frac{1-2G\Pi_M(\omega-i\eta,\vec q)}
{1-2G\Pi_M(\omega+i\eta,\vec q)}\pkt
\eeq
The resulting mesonic contributions to the pressure in the three-flavor PNJL model are displayed in 
Fig. \ref{fig:pfluct} \cite{Radza}. 
As mentioned above, the thermal quark- and diquark excitations are supressed by the Polyakov loop in the confined phase and the pressure comes entirely from mesons, as it should be. Up to temperatures of about 150 MeV it agrees well with the model-independent results from chiral perturbation theory \cite{GaLeu}. When quarks and gluons take over above 200 MeV, the mesonic contributions become supressed, since the meson masses grow rapidly, reaching the thermal value of $2\pi T$ around 300 MeV. Hence they become strongly Boltzmann-supressed and the 'Hagedorn singularity' is avoided.

Now is is easy to see why the chiral quark-hadron transition is softened when including fluctuations beyond the mean field. The chiral condensate is given by the derivative of the pressure (or $\Omega$)
with respect to the bare quark mass: $\ave{\bar qq}=\partial\Omega/\partial m_q$. Hence in the confined phase 
\beq
\ave{\bar qq}=\frac{\partial\delta\Omega}{\partial m_q}=\sum_M\frac{\partial\Omega_M}{\partial M}\frac{\partial M}{\partial m_q}
\eeq
which is non-vanishing. Since the Polyakov loop dynamically couples to the quaks its softening can be understood by the same token. 

\section{'Quarkyonic' Phase at low $T$ and large $\mu$?}
\label{sec:2}

Based on large-$N_c$ QCD it has been argued that there should exist a phase of strong-interaction matter in which the chiral and the deconfined transition split apart, since the deconfinement transition temperature becomes independent of the quark chemical potential \cite{LePi}. Within a narrow window in $\mu$ with a width of the order of $\sim 1/N_c^2$ there should then be a rapid transition to a dense phase of confined hadrons (mostly baryons at low $T$) in which chiral symmetry is restored leading to parity doubled color singlet excitations. Schematically this leads to the following phase diagram:
\begin{figure}[h]
\resizebox{0.8\columnwidth}{!}{%
\hspace{0.5cm}\includegraphics{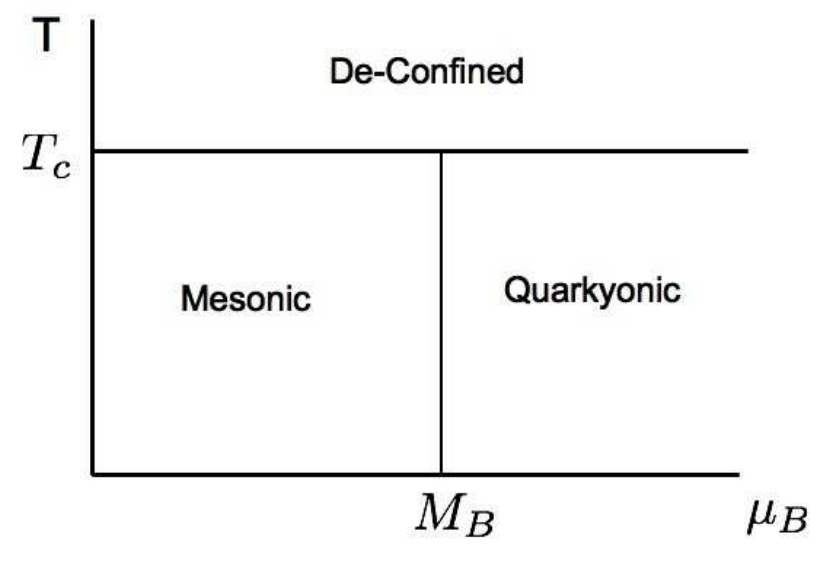}}
\caption{Sketch of the large-$N_c$ QCD phase diagram as it emerges from the arguments of Ref. 
\cite{LePi}. All phase boundaries are of first order with a 'triple point' where all first order lines meet. As a function of $\mu_B$ the transition happens at the vacuum baryon mass $M_B\sim N_c\Lambda_{QCD}$. }
\label{fig:Quarkyonic}       
\end{figure}

\begin{figure}[h]
\resizebox{1.0\columnwidth}{!}{%
\includegraphics{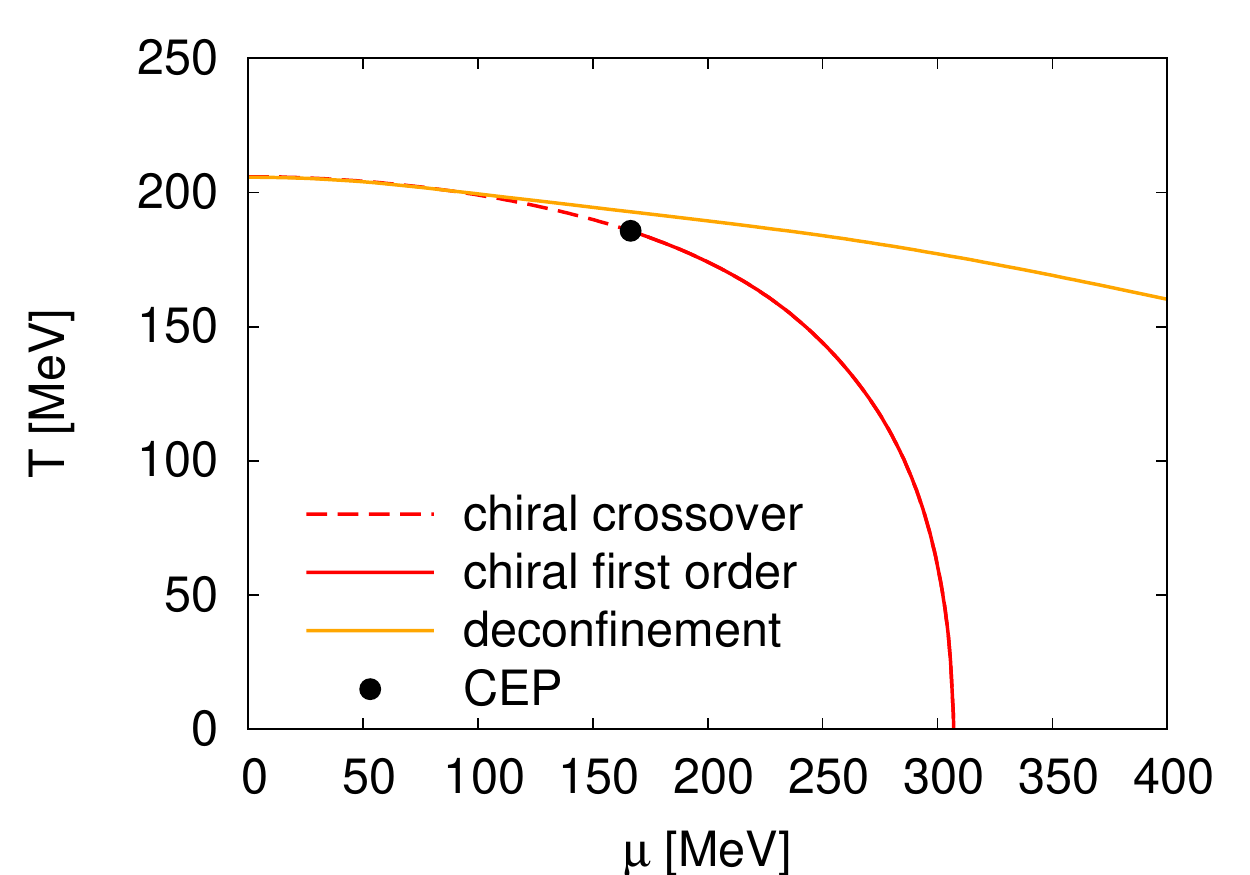}}
\caption{Mean-field results for the phase diagram in the $N_f=3$ PQM model \cite{SWW}. The deconfinement boundary is determined by the inflection point in the Polyakov-loop expectation 
value $\ell$.}
\label{fig:Nc=3}      
\end{figure}

Within a PNJL model for $N_c\to\infty$ and a fixed number of flavors this picture can be reproduced in the mean-field aproximation \cite{Sasaki}. At $N_c=3$ the situation changes, however. Fig. \ref{fig:Nc=3} showns a PQM calculation of the phase diagram for three colors and three flavors \cite{SWW}. 
In the physical word the (pseudo)critical transition lines curve downwards with increasing $\mu$ and the  triple point evolves into the CEP.   As one can see, there is a large region in which there still is confinement (in the statistical sence) but chiral symmetry is restored which thus qualifies as the quarkyonic phase. The size of this phase, however, crucially depends on the Polyakov-loop dynamics. 

Let us consider the polynomial ansatz for the Polyakov-loop potential for example
\beq
\frac{{\cal U}(\ell,\bar\ell)}{T^4}=-\frac{b_2(T)}{2}\ell\bar\ell-\frac{b_3}{6}\left(\ell^3+\bar\ell^3\right)
+\frac{b_4}{16}\left(\ell\bar\ell\right)^2
\eeq
with
\beq
b_2(T)=a_0+a_1(T_0/T)+a_2(T_0/T)^2+a_3(T_0/T)^3\pkt
\eeq
Originally the parmeter $T_0$ was adjusted to the first-order deconfinement transition temperature 
of $T=270$ MeV in 'pure gauge' theory, i.e. without dynamical quarks. This, however, does not take into account that with dynamical quarks $T_0$ acquires a $N_f$- and $\mu$ dependence. Based on the one-loop running of the QCD $\beta$-function this dependence has been estimated in Ref. \cite{SPW} as
\beq
T_0(N_f,\mu)=T_\tau \exp\left(-1/\alpha_0b(N_f,\mu)\right)
\label{eq:T0} 
\eeq
with
\beq
b(N_f,\mu)=\frac{1}{6\pi}(11N_c-2N_f)-\frac{16 N_f}{\pi}\frac{\mu^2}{T_\tau^2}
\eeq
where $\alpha_0=\alpha(\Lambda)$ is the running gauge coupling at some $UV$-scale and 
$T_\tau=1.777$ GeV is fixed at the scale of the $\tau$-meson to reproduce $T_0=270$ MeV at $N_f=0$ with the corresponding value $\alpha_0=0.304$. 

\begin{figure}[h]
\resizebox{0.9\columnwidth}{!}{%
\includegraphics{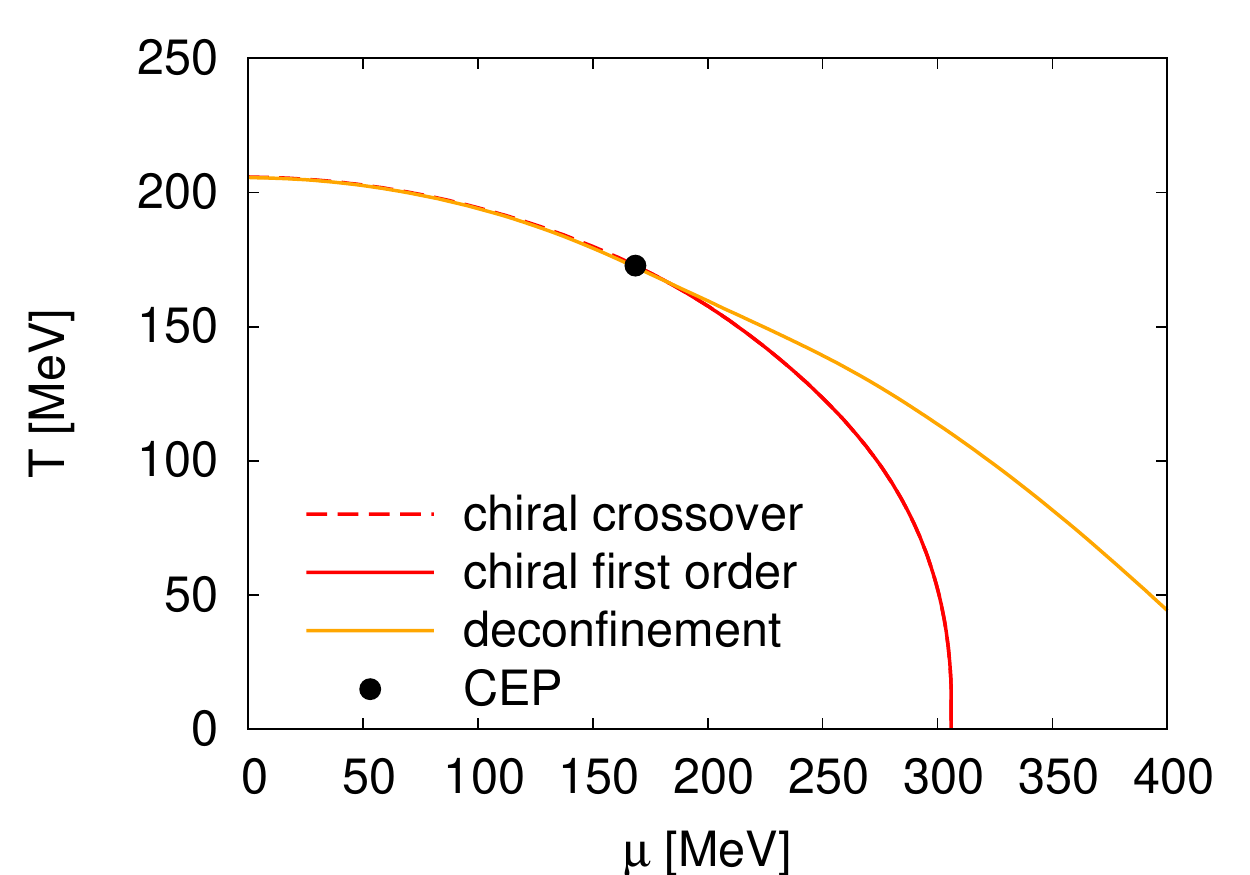}}
\resizebox{0.9\columnwidth}{!}{%
\includegraphics{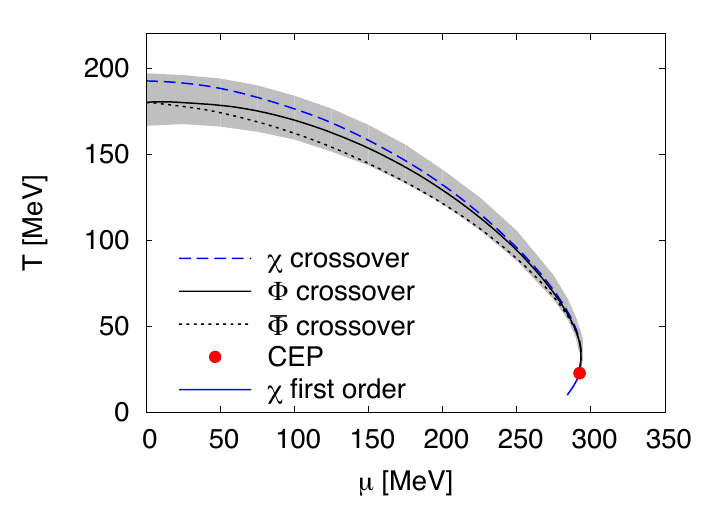}}
\caption{Upper part: mean-field results for the phase diagram in the $N_f=3$ PQM model \cite{SWW} with $T_0(N_f,\mu)$. Lower part: FRG result for the $N_f=2$ PQM model \cite{Herbst}  .}
\label{fig:T0mu}      
\end{figure}

At $\mu=0$ this leads to the Polyakov-loop $T_0$ given in Tab. 1. Of importance for the size of the quarkyonic phase is the $\mu$-dependence of $T_0$.  
\begin{table}[h]
\resizebox{1.0\columnwidth}{!}{%
\includegraphics{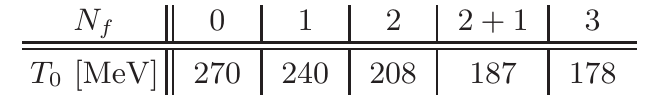}}
\caption{$T$ and $\mu$ dependence of $T_0$ \cite{SPW}.}
\end{table}
When using the estimate in Eq. (\ref{eq:T0}) one obtains in a PQM calculation instead of Fig. \ref{fig:Nc=3} the result displayed in the upper part of Fig. \ref{fig:T0mu}.  
The region of the quarkyonic phase has considerably shrunk. This situation is even more dramatic if, instead of a mean-field calculation, one includes quantum fluctuations via the FRG \cite{Herbst} (lower part of Fig. \ref{fig:T0mu}). In this case the chiral and the deconfinement transition essentially coincide and hence the region of quarkyonic matter has collapsed to nearly zero.

\begin{figure}[h]
\resizebox{0.9\columnwidth}{!}{%
\includegraphics{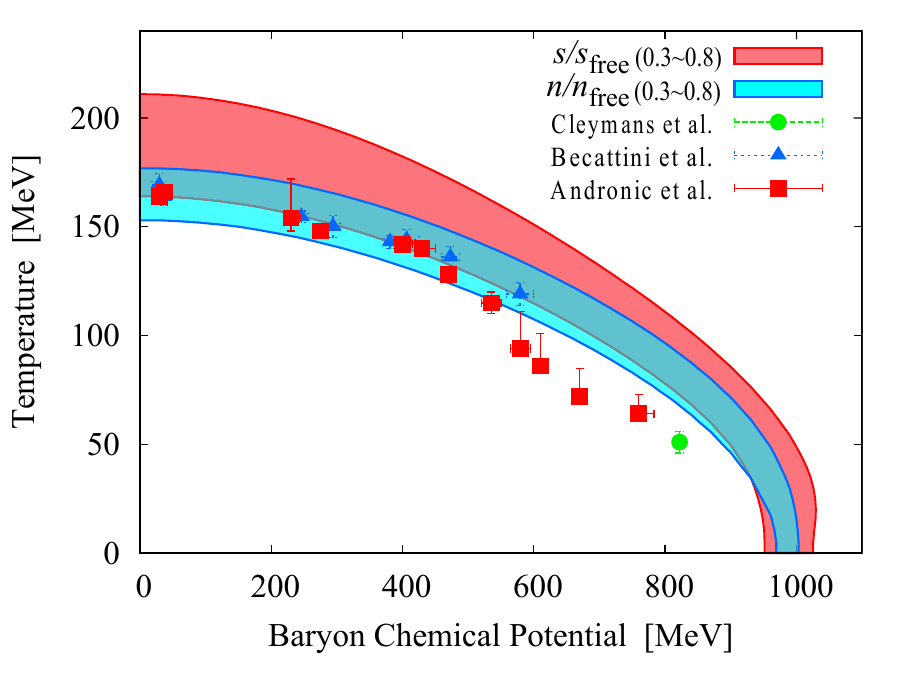}}
\resizebox{0.9\columnwidth}{!}{%
\includegraphics{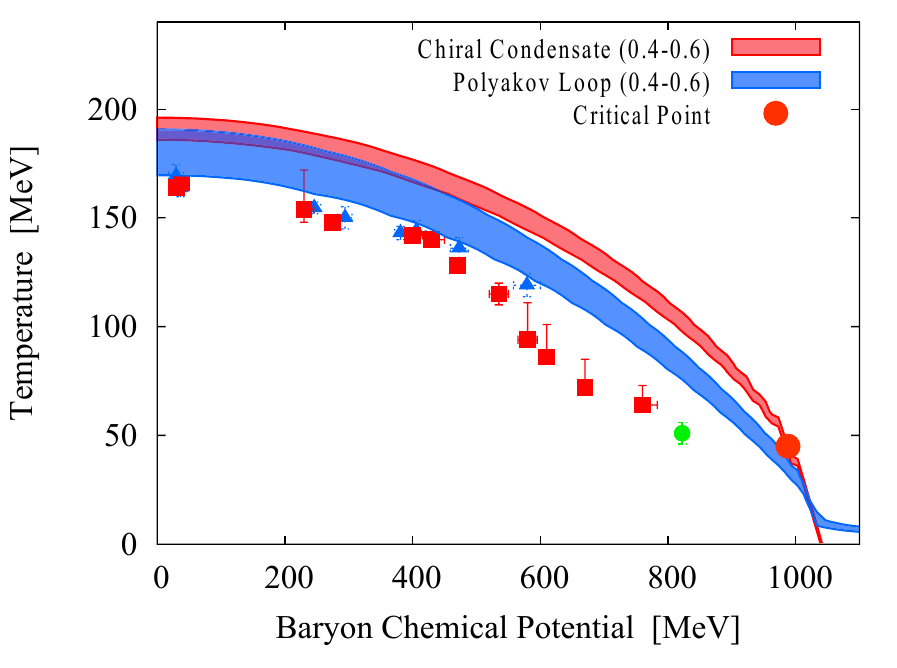}}
\caption{Upper part: The normalized entropy- and baryon number density of the hadron resonance gas as a function of $T$ and $\mu$ \cite{Fuku-RG}. The red and blue bands indicate the regions, where both quantities increase from 0.3-0.8. The chemical freeze-out data are taken from \cite{Cleymans,Becattini}. Lower part: boundaries of the regions in which the chiral condensate and the Polyakov loop change between 0.4 and 0.6.}
\label{fig:StatMod}      
\end{figure}

There is a more heuristic argument, why the boundaries for the chiral- and deconfinement transitions 
cannot be to far appart \cite{Fuku-RG}. This is based on the statistical model of a hadron resonance gas which is extremely successful in describing the chemical freeze out of hadrons in relativistic heavy-ion collisions over a very wide range of beam energies\cite{PBMW}. The thermodynamics of the resonance gas is that of a free gas and therefore the EoS can be straightforwardly evaluated. Results for the entropy density $s$ and the baryon number density $n$, normalized to the free quark-gluon gas values are displayed in the upper part of Fig. \ref{fig:StatMod}. 

\noindent
The idea is now to take the $\mu$ dependence of the $T_0$ parameter in the Polyakov-loop potential ${\cal U}$ directly from the $\mu$ dependence of the empirical freeze-out line. With 
\beq  
T_0(\mu)/T_0(0)=1-(bT_0)(\mu_B/T_0)^2
\eeq
and fixing $T_0$ at 200 MeV this yields $bT_0=2.78\times 10^{-2}$ MeV, in quite good agreement with   the estimates of Ref. \cite{SPW} which give $T_0=187$ MeV and $bT_0=2.1\times 10^{-2}$ MeV. The mean-field PNJL calculation with this parametrization of $T_0(\mu)$ reproduces the entropy- and number density values of the resoncance gas, shown in the upper part of Fig. \ref{fig:StatMod}, very well. The $T$ and $\mu$ dependence of the chiral condensate and the Polyakov loop can then be inferred from the calculation. The results are shown in the lower part of Fig. \ref{fig:StatMod} and again indicate that the chiral- and deconfinement transition line are close in the whole $(T,\mu)$-plane, leaving little room for a quarkyonic phase.     

\section{Inhomogeneous Phases of QCD Matter}
\label{sec:3}

In the conventional picture which emerges from model calculations of the PNJL 
or PQM type, the chiral transition in the $(T,\mu)$-plane is a smooth cross over at low
$\mu$ and large $T$ (see Sect. \ref{sec:1}). It becomes a first-order 
phase transition at large $\mu$ and small $T$ from the chirally broken to the chirally 
restored phase. Usually it is assumed that both phases are homogeneous. From studies
of 1+1 dimensional fermionic theories, such as the Gross-Neuveu, the NJL or the 
t'Hooft model it is known analytically that in some regions of the $(T,\mu)$-plane, 
inhomogeneous phases are preferred in the $N\to\infty$ limit \cite{Schnetz}. When applying 
these results in a Ginz- burg-Landau analysis near the CEP (which is a second-order phase
transition) it is found that the homogeneous phase in three space dimensions is unstable
against one-dimensional spatial oscillations of the chiral order parameter \cite{Nickel-LG}.   

\subsection{Phase diagram in 1+1 dimensional theories} 

Let us consider the Gross-Neveu (GN) model with $N$ degrees of freedom as an example. 
With strong-interaction physics it shares some essential features, such as asmptotic freedom, 
dimensional transmutation and spontaneous chiral symmetry breaking in the vacuum.  
The GN model is specified by the following Lagrangian:    
\beq
\mathcal{L}_{GN} =
\sum_{i=1}^N\bar{\psi}^{(i)}\left(i\dslash-m_0\right)\psi^{(i)}+
\frac{g^2}{2}\left(\sum_{i=1}^N\bar{\psi}^{(i)}\psi^{(i)}\right)^2
\eeq
where $\psi$ is an $N$-component fermion field, $g^2$ the coupling constant of the four-fermion interaction and $m_0$ the bare fermion mass. The model posesses a global $U(N)$ symmetry 
and a $Z_2$ chiral symmerty:
\beq
\psi\to\gamma_5\psi,\quad \bar\psi\psi\to-\bar\psi\psi\pkt
\eeq
As usual, mean field theory becomes exact in the $N\to\infty, Ng^2=const$ limit and the resulting 
selfconsistent Hartree equations for given $T$ and $\mu$ read:  
\bea
\left(-i\gamma_5\partial_z+\gamma^0M(z)\right)\psi_\alpha=\epsilon_\alpha\psi_\alpha
\nonumber\\
M-m_0=-Ng^2\sum_\alpha n_\alpha\bar\psi_\alpha\psi_\alpha\pkt
\eea
Here $n_\alpha(T,\mu)$ denote the Fermi-Dirac occupation probabilities. The mean-field equations can be solved analytically with the $z$-dependent (constituent) mass function given by an elliptic 
Jakobi-sn function in the chiral limit, $m_0=0$:  
\beq
M(z)=\sqrt{\nu}q\,{\rm sn}(qz|\nu)\pkt
\eeq
The elliptic modulus $\nu$ varies continuously between zero and one and $q$ is a scale, related to the maximum of $M(z)$. For $\nu=1$ one has $M(z)=q\tanh(qx)$, i.e. a single soliton and for $\nu\to0$ the shape becomes more and more sinusoidal, albeit the amplitude also goes to zero. Thus the sn-function interpolates smoothly between soliton-like and sinusoidal shapes.
   
\begin{figure}[h]
\resizebox{1.0\columnwidth}{!}{%
\includegraphics{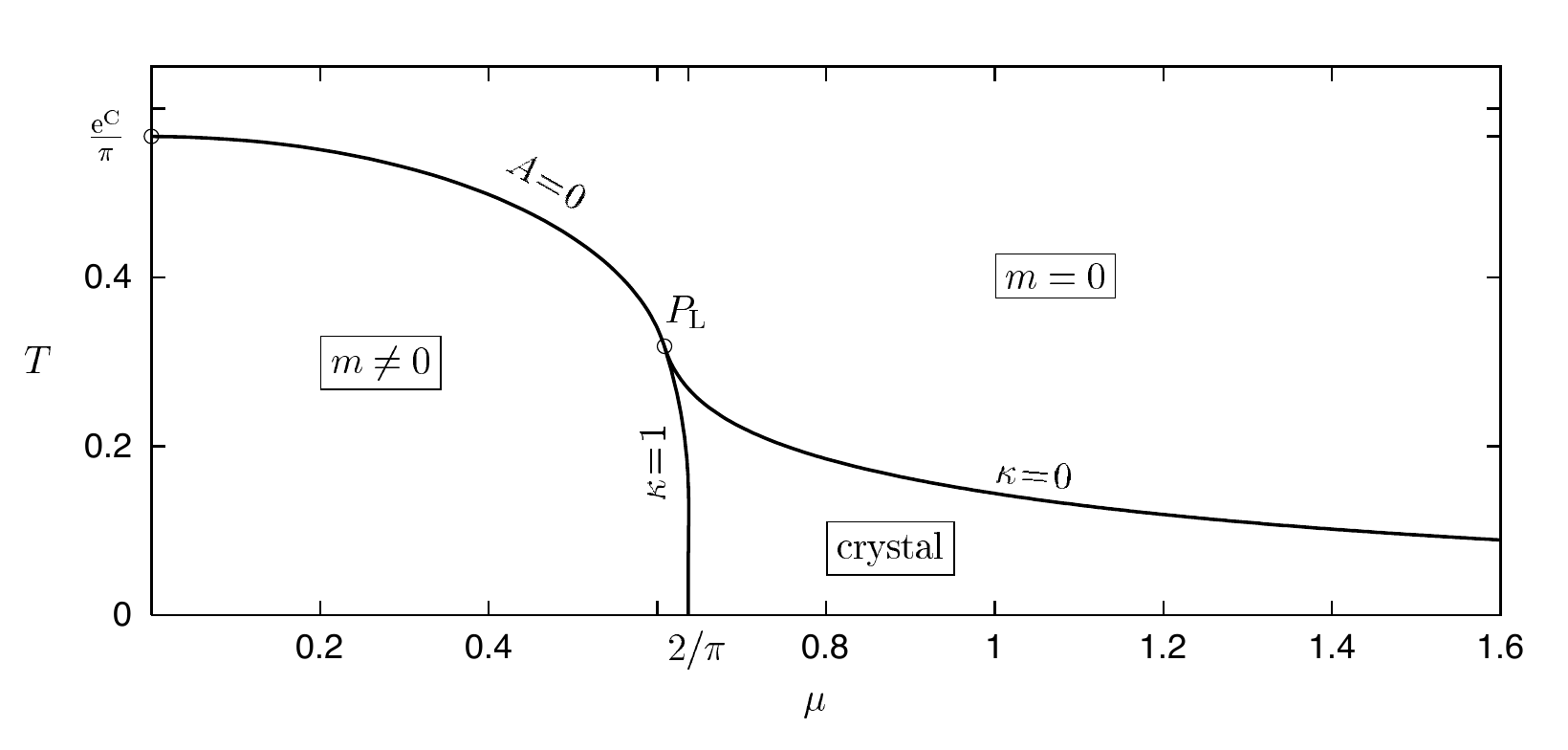}}
\resizebox{1.0\columnwidth}{!}{%
\includegraphics{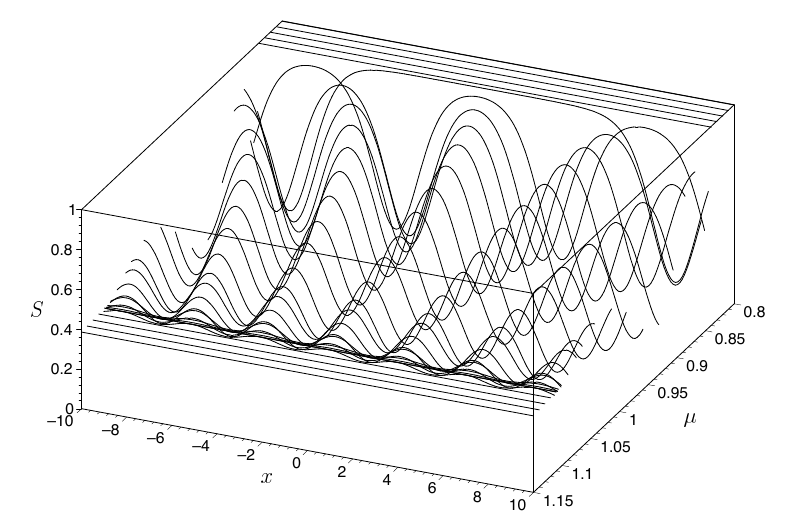}}
\caption{Upper part: phase diagram of the exactly solvable $1+1$ dimensional Gross-Neveu model
in the large-$N$ limit \cite{Schnetz}. A spatially ordered phase occurs for small temperatures and $\mu>2/\pi$. The phase boundaries deliniate second-order transitions which meet at a tricritical point, the 'Lifschitz' point. Lower part: evolution of the spatial modulations of the mass function $M(z)$ as one moves along a line of constant (small) $T$. While at smaller $\mu$ the modulation is solitonic it becomes increasingly sinusoidal with increasing $\mu$.   
}
\label{fig:PD-1D}      
\end{figure}

The actual values for given $T$ and $\mu$ are determined from the minimization of the grand potential 
\beq
\Omega(T,\mu)=-T\,\mathrm{Tr}\, \mathrm{log}\left(S^{-1})\right)
+\inv{2Ng^2\lambda}\int_0^\lambda\!\! dz\;M(z)^2
\eeq
where $S$ is the fermion propagator in the Hartree approximation and $\lambda$ the period of the spatial modulation. The resulting phase diagram in the chiral limit is shown in in the upper part of Fig. \ref{fig:PD-1D}. 
Raising $T$ for $\mu< 2/\pi$ one encounters a line of second order transitions from the chirally 
broken, $M\neq 0$, to the chirally restored phase, $M=0$, both of them spatially homogeneous.
Cutting through the phase diagram at small constant $T$, one enters a region of inhomogeneous phases through a second-order transition, where the spatial modulation go from soliton-like to sinusoidal as $\mu$ increases (lower part of Fig. \ref{fig:PD-1D}). At large enough $\mu$ and moderate $T$ one again enters the homogeneous chirally restored phase through a second order transition. All lines of second order transitions meet in a tricritical point, which is commonly referred to as the 'Lifschitz' point.

\subsection{Inhomogeneous phases in the NJL model}

As mentioned above, QCD-inspired models of the NJL or QM type are currently used to 
asses details of the phase diagram at large chemical potentials, where
ab-initio lattice methods fail. It is therefore interesting to ask whether spatially inhomogeneous 
regions also occur in three spatial dimensions and what their properties are. This 
question has been addressed recently for 1D modulations of the chiral order parameter 
in 3D space (plates) \cite{Nickel-NJL}. 

The starting point of the analysis is the two-flavor NJL model with scalar coupling:       
\beq
\mathcal{L}_{NJL} =\bar{q}\left(i\dslash-m_q\right)q+
G_s\left(\left(\bar{q}q\right)^2+\left(\bar{q}i\gamma^5\tau^aq\right)^2\right)\pkt
\eeq
Considering phases with a spatially varying expectation values $\ave{\bar qq(\bf x)}=S(\bf x)$
and $\ave{\bar qi\gamma^5\tau^aq(\bf x)}=P_a(\bf x)$ and restricitng to the case where the direction 
of the vector $P_a(\bf x)$ is constant in flavor space such that $P_1({\bf x})=P_2({\bf x})=0$ and 
$P_3({\bf x})=P({\bf x})$, the following mean-field Lagragian is obtained:
\bea
{\cal L}_{NJL}^0({\bf x})&=&\bar q\left(i\dslash-m_q+2G_s\left(S({\bf x})
+i\gamma^5\tau_3P({\bf x})\right)\right)q\nonumber\\
&&-G_s\left(S({\bf x})^2+P({\bf x})^2\right)\pkt
\eea
In terms of the (complex) mass function
\beq
M({\bf x})=m_q-2G_s\left(S({\bf x})+iP({\bf x})\right)
\eeq
the mean-field grand potential for 1D modulations can be evaluated straightforwardly and one obtains 
\bea
\Omega(T,\mu)&=&-\frac{2T}{V}\sum_\alpha\!\!\int_{p_\perp}\!\!\!\frac{d^2{\bf p}_\perp}{(2\pi)^2}\ln\left(2\cosh\left(
\frac{\alpha\sqrt{1+{\bf p}^2_\perp/\alpha^2}-\mu}{2T}\right)\right)
\nonumber\\
&&+
\int_V\frac{\vert M({\bf x})-m_q\vert^2}{4G_s V}
\eea
where $V$ is the volume of the Wigner-Seitz cell of the periodic condensate. Here it is assumed that the 
inhomogeneity is in the $z$-direction and the perpendicular $(x,y)$-plane is translationally invariant. The discrete parameter $\alpha$ involves the Eigenvalues of the mean-field Hamiltonian in the $z$-direction which can be obtained analytically for the case in which $M(\bf x)$ is real\footnote{A complex mass function $M({\bf x})$ leads to so-called 'chiral spirals' for sinusoidal modulations \cite{Basar}}. It can be argued that a real order parameter is thermodynamically preferred, at least in the vicinity of a second-order transition and in the chiral limit \cite{Nickel-LG}.

\begin{figure}[h]
\resizebox{1.0\columnwidth}{!}{%
\includegraphics{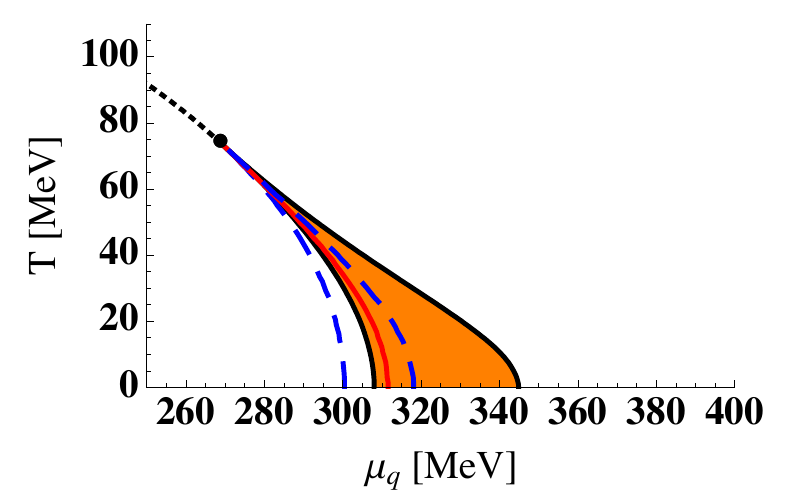}}
\resizebox{1.01\columnwidth}{!}{%
\includegraphics{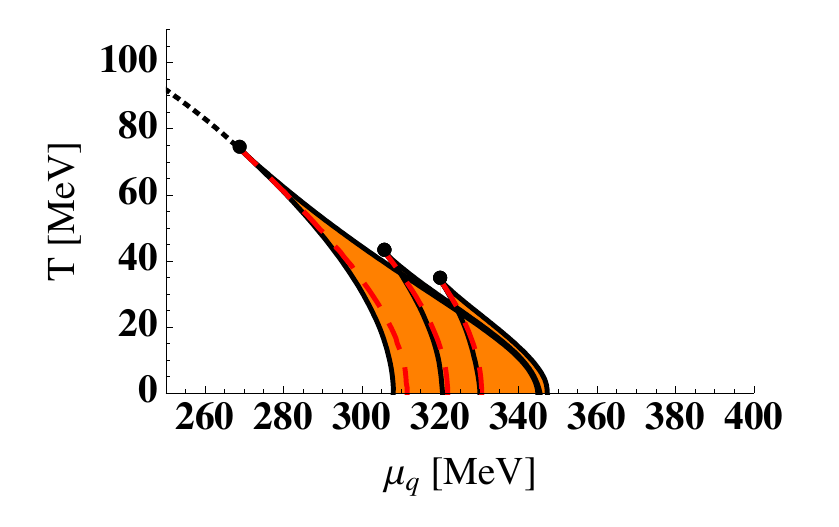}}
\caption{Upper part: phase diagram of the $N_f=2$ NJL model in the chiral limit \cite{Nickel-NJL}. The shaded area denotes the region of 1D plates in the $z$-direction. The (tricritical) CEP of the homogeneous phases coincides with the Lifschitz point. The dashed lines denote the spinodal lines of the homogeneous phases.
Lower part: same as the upper part, but now including finite bare quark masses of $m_q=5$ MeV and 10 MeV \cite{Nickel-NJL}.   
}
\label{fig:PD-1DNJL}      
\end{figure}

As can be seen from Fig. \ref{fig:PD-1DNJL} the homogeneous first-order chiral transition is completely covered by an inhomogeneous region, bounded by second-order transition lines. As one goes away from the chiral limit (lower part of Fig. \ref{fig:PD-1DNJL}) the Lifschitz point moves to higher $\mu$ and lower $T$ and the inhomogeneous region shrinks\footnote{The same conclusions also hold in the QM model 
\cite{Nickel-NJL}.}.     
The phase diagram in the 3+1 dimensional NJL model looks very similar to the 1+1 case of the GN model, except that the inhomogeneous region does not extent to large $\mu$. This may be due to cut-off effects \cite{Buballa}. While the 1+1 dimensional GN model is renormalizable, this is not the case for the NJL model in three dimensions. Here the loop intergrals have to regularized and the ultraviolet cut-off enters as an explicit parameter into the calculation.   

\subsection{Including Vector Interactions}

It is well known that extended (P)NJL models which include vector interactions lead to substantial 
modifications of the location of the CEP in the phase diargram. For sufficiently large vector-interaction 
strength, the CEP even disappears and the chiral transition becomes a cross over in the whole 
$(T,\mu)$ plane. In view of the above discussion is therefore important to ask, how spatial inhomogeneities of the chiral order parameter influence these findings.

\begin{figure}[h]
\resizebox{1.0\columnwidth}{!}{%
\includegraphics{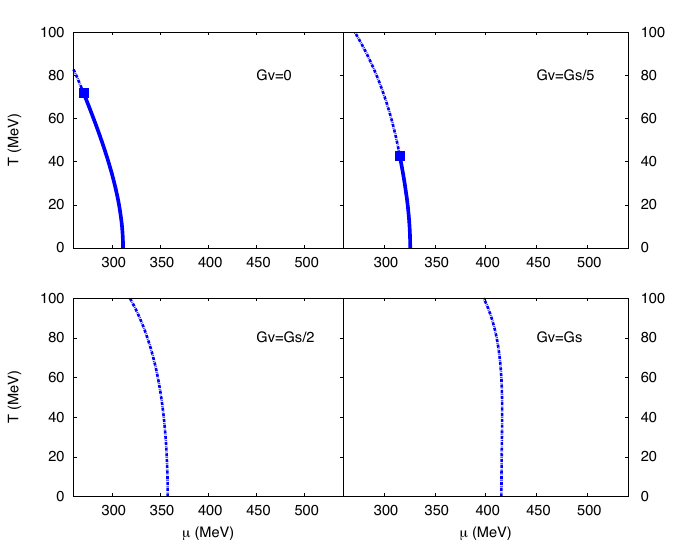}}
\resizebox{1.0\columnwidth}{!}{%
\includegraphics{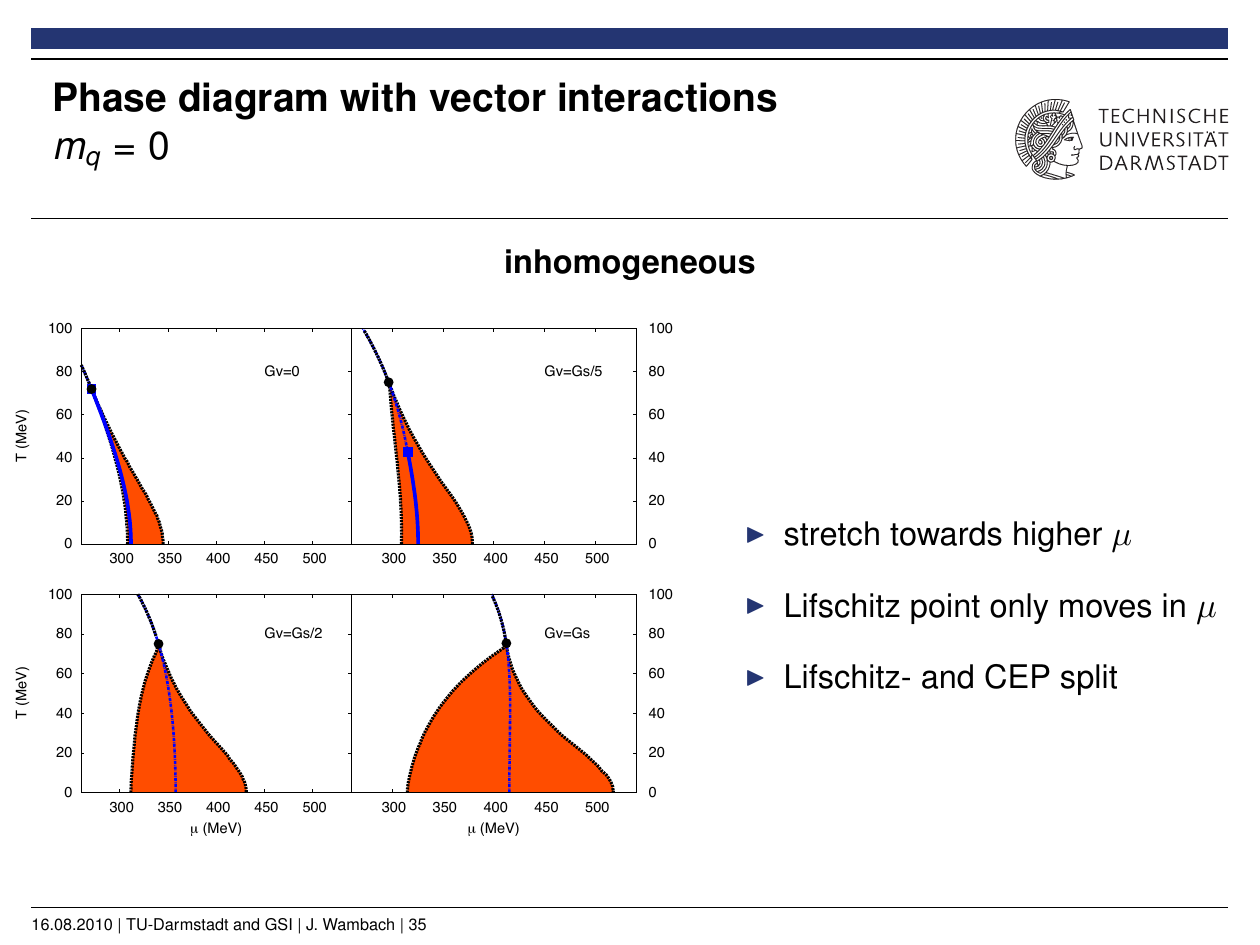}}
\caption{Upper part: The homogeneous phase diagram including vector interactions of increasing strength (left to right). At sufficiently large coupling the CEP disappears. Lower part: same but allowing for 1D inhomogeneities \cite{Cari1}.}
\label{fig:PD-HNJLV}      
\end{figure}

Including vector interactions the $N_f=2$ NJL Lagrangian reads
\beq
\mathcal{L}_{NJL} =\bar{q}\left(i\dslash-m_q\right)q+
G_s\left(\left(\bar{q}q\right)^2+\left(\bar{q}i\gamma^5\tau^aq\right)^2\right)-G_V(\bar q\gamma^\mu q)^2\pkt
\eeq
Varying $G_V$ strongly affects the location of the CEP when only homogeneous phases are considered (upper part of Fig. \ref{fig:PD-HNJLV}).  

One-dimensional spatial inhomogeneities can be treated in a similar way as for the NJL model with scalar interaction only. An additional complication arises, however, from the fact that the vector interaction couples to the density which is also spatially modulated, i.e. $n=\ave{\bar q\gamma^0 q}= n(\bf x)$. Via
the renormalized quark chemical potential
\beq
\tilde\mu({\bf x})=\mu-2G_Vn({\bf x})
\eeq
this also induces a spatially varying chemical potential. This complicates the minimization of the grand potential considerably. In order to take advantage of the fact that, for 1D modulations, the partition sum of the NJL model without vector interaction can be performed analytically for states in the $z$-direction, one has to replace density by its spatial average
\beq
n({\bf x})\to \bar n\equiv \ave{n({\bf x})}={\rm const.}
\eeq
which is only approximate, but can be justified rigourously in the vicinity of a second-order boundary to the restored phase and in particular near the Lifschitz point \cite{Cari1}. As a consequence $\tilde\mu$ becomes constant as well and the problem reduces to the case discussed in the previous Sect., albeit with shifted chemical potential $\tilde\mu$. Another interesting consequence of non-vanishing $G_V$ is that the CEP and Lifschitz point no longer coincide. As shown in the lower part of Fig. \ref{fig:PD-HNJLV}, increasing $G_V$ leads to two effects: (1) the CEP moves deeper into the inhomogeneous region and eventually dissappears, (2)
the inhomogeneous region grows . The latter is largely a trivial effect of the shifted chemical potential. Plotting the phase diagram in terms of the average density, instead of $\tilde\mu$, the transition lines do not depend on $G_V$ at all. It is remarkable that, although the CEP is strongly effected by the vector interaction, the location of the Lifschitz point is not (except for a more or less trivial $\mu$-dependence). This speaks for the 'robustness' of the inhomoneneous region.

The influence of gluonic degrees of freedom on the occurence of spatially modulated phases via the coupling to the Polyakov loop has also been considered in Ref. \cite{Cari1}. Preliminary results indicate that the general picture remains unchanged.   

It has long been known that 1D long-range order is thermodynamically unstable 
\cite{Peierls,Landau}. In the context of pion-condensed phases this has been reiterated by Baym et al.
\cite{Baym}. The occurence of 1D order is a mean-field or large-$N$ artefact and any finite temperature destroys the long-range order. There remains however the possibilty of a quasi-ordered one-dimensional phase, with long-range correlations decaying algebraically in space \cite{Baym}. In any case, two- and three-dimensional structures will be stable and it is worth exploring them \cite{Cari2}.       

\section{Conlusions and Outlook}
\label{sec:4}

Much progress has been made in recent years in the theoretical exploration of the phase diagram of hadronic matter. Largely based on calculations in QCD-inspired models a rich structure has emerged, especially at large chemical potentials and low temperatures. With the help of FRG methods is has become possible the include fluctuations on top of the mean-field results \cite{SWa,SSFR,Herbst}. These are able to properly describe the critical exponents at the CEP \cite{SWa}, lead to a small critical region around the CEP and result in a softening of chiral and deconfinement transition as compared to mean-field predictions at small $\mu$. The latter finding is consistent with state-of-the-art lattice simulations of the quark-hadron transition at vanishing $\mu$ and has a simple physical interpretation in terms of meson fluctuations below the pseudo-critical temperature \cite{Radza}. 

An interesting question that has caught much attention recently is the existence of a quarkyonic phase which is confining but chirally restored. The arguments for such a phase are based on a large $N_c$ analysis of the pressure at finite $T$ and $\mu$. For $N_c=3$ there are substantial modifications and it is not clear in what form the large-$N_c$ picture survives. Mean-field PNL and PQM calculations indicate that there is a region in the phase diagram in which quarkyonic matter could exist. The size of this region crucially depends of the parameters of the Polyakov-loop potential, in particular the $\mu$-dependence of $T_0$.
Present FRG calculations indicate that the deconfinement and chiral transition lines in the $(T,\mu)$ plane almost coincide \cite{Herbst}, leaving little room for a quarkyonic matter state. These findings are corroborated by a heuristic analysis based on the resonance-gas EoS and the experimental location of the chemical freeze-out points in the phase diagram \cite{Fuku-RG}.

An exciting new possibility is the occurence of spatial modulations of the chiral order parameter, leading
to a ordered density profile. Exact 1+1 dimensional results for QCD-like models in the 
large-$N_c$ limit can be taken over to three space dimensions in the PNJL and the PQM model \cite{Nickel-NJL} and lead to plate-like structures. The inhomogeneous phases are bounded by second-order transition lines and lead to a Lifschitz point in the chirial limit. Without vector interactions, the Lifschitz point coincides with the CEP but both drift appart with increasing vector coupling \cite{Cari1}. One can even achieve scenarios in which the CEP dissapears while the inhomogeneous phase region remains. It is very robust and its size is unaffected when considering the average quark-number density rather that $\mu$. One-dimensional phase are thermodynamically unstable although quasi one-dimensional structures ar still possible. Two- and three dimensional spacial ordering is likely to occur and should be studied \cite{Cari2}. It is well established theoretically that such phases are posible near the crust-liquid interface in a neutron star and it would be very exciting if they would also show up in the quark-hadron transition in its inner core.
\\\\

\noindent    
{\bf Acknowledgement}\\

\noindent
I thank my collaborators for numerous discussion. This work has been supported in part by BMBF grant 06DA9047I, the Helmholtz Alliance EMMI and the Helmholtz International Center for FAIR.


\begin{thebibliography}{99}

\bibitem{PBMW}
P. Braun-Munzinger and J. Wambach, Rev. Mod. Phys. {\bf 80} (2009) 1031.

\bibitem{Alford} 
M.G. Alford, A. Schmitt, K. Rajagopal and T. Schaefer, Rev. Mod. Phys. {\bf 80} (2008) 1455.

\bibitem{Ratti} 
C. Ratti, M.A. Thaler and W. Weise, Phys. Rev. {\bf D73 } (2006) 014019.

\bibitem{Roesner}
S. Roessner, C. Ratti and W. Weise, Phys. Rev. {\bf D75} (2007) 034007.

\bibitem{FukuPot}
K. Fukushima, Phys. Rev. D {\bf D77} (2008) 114028.

\bibitem{SWW}
B.-J. Schaefer, M. Wagner and J. Wambach, Phys. Rev. {\bf D81} (2010) 074013.

\bibitem{HOTQCD}  M. Cheng et al. Phys. Rev. {\bf D74} (2006) 054507,\\ 
A. Bazavov et al., Phys. Rev. {\bf  D80} (2009) 014504.  

\bibitem{WB} Y. Aoki et al., Phys. Lett. {\bf B643} (2006) 46.

\bibitem{SWa}
B.-J. Schaefer and J. Wambach, Nuch. Phys. {\bf A757} (2005) 479.  

\bibitem{SSFR}
V. Skokov et al., Phys. Rev. {\bf C82} (2010) 015206. 

\bibitem{Oertel}
M. Oertel, M. Buballa and J. Wambach, Nucl.Phys. {\bf A676} (2000) 247.

\bibitem{Radza}
A. Radzhabov et al., arXiv:1012.0664 [hep-ph].

\bibitem{GaLeu}
J. Gasser and H. Leutwyler, Phys. Lett. {\bf B184} (1987) 83.

\bibitem{LePi}
L. McLerran and R. Pisarski, Nucl. Phys. {\bf A796} (2007) 83.

\bibitem{Sasaki}
L. McLerran, K. Redlich, C.Sasaki, Nucl. Phys. {\bf A824} (2009) 86.

\bibitem{SPW}
B.-J. Schafer, J. Pawlowski and J. Wambach, Phys. Rev. \textbf{D76}, (2007) 074023.

\bibitem{Herbst}
T.K. Herbst, J. Pawlowski and B.-J. Schaefer, Phys. Lett. {\bf B696} (2011) 58.

\bibitem{Fuku-RG}
K. Fukushima,  arXiv:1006.2596 [hep-ph]. 

\bibitem{Cleymans}
J. Cleymans et al., Phys. Rev. {\bf 73} (2006) 034905.

\bibitem{Becattini}
F. Becattini, J. Manninen and M. Gazdzicki, Phys. Rev. {\bf C73} (2006) 044905,\\
A. Andronic, P. Braun-Munzinger and J. Stachel, Phys. Lett. {\bf B673} (2009) 142.

\bibitem{Schnetz}
O. Schnetz, M. Thies and K. Ulrichs, Annals Phys. {\bf 314} (2004) 425,\\
O. Schnetz, M. Thies and K. Ulrichs, Annals Phys. {\bf 321} (2006) 2604.

\bibitem{Nickel-LG}
D. Nickel, Phys. Rev. Lett.{\bf 103} (2009) 072301.

\bibitem{Nickel-NJL}
D. Nickel, Phys. Rev. {\bf D80} (2009) 074025.

\bibitem{Buballa}
M. Buballa, private communication. 

\bibitem{Basar}
G. Basar, G.V. Dunne and M. Thies, Phys. Rev. {\bf D79} (2009) 105012.  

\bibitem{Peierls}
R.E. Peierls, Helv. Phys. Acta Suppl. {\bf 7} (1934) 81.

\bibitem{Landau}
L.D. Landau and E.M. Lifshitz, Statisitical Physics (Addison Wesley, Reading, Mass., 1969) 402.

\bibitem{Baym}
G. Baym, B. Friman and G. Grinstein, Nucl. Phys. B {\bf 210} (1982) 193.

\bibitem{Cari1}
S. Carignano, D. Nickel and M. Buballa, Phys.Rev. D {\bf 82} (2010) 054009. 

\bibitem{Cari2}
S. Carignano, M. Buballa and J. Wambach, work in progress.

\end{thebibliography}
\end{document}